\documentclass[11pt]{article}
\usepackage{amsmath,amssymb,color}

\usepackage{amsfonts}

\newcommand{\hoch}[1]{$\, ^{#1}$}

\newcommand{\be}{\begin{equation}}
\newcommand{\ee}{\end{equation}}
\newcommand{\bea}{\setlength\arraycolsep{2pt} \begin{eqnarray}}
\newcommand{\eea}{\end{eqnarray}}
\newcommand{\nn}{\nonumber}

\def\ft#1#2{{\textstyle{\frac{\scriptstyle #1}{\scriptstyle #2} } }}
\def\fft#1#2{{\frac{#1}{#2}}}

\def\0{{\sst{(0)}}}
\def\1{{\sst{(1)}}}
\def\2{{\sst{(2)}}}
\def\3{{\sst{(3)}}}
\def\4{{\sst{(4)}}}
\def\5{{\sst{(5)}}}
\def\6{{\sst{(6)}}}
\def\7{{\sst{(7)}}}
\def\8{{\sst{(8)}}}
\def\sst#1{{\scriptscriptstyle #1}}

\def\del{{\partial}}

\def\cG{{{\cal G}}}

\begin{document}

\begin{flushright}
\hfill{ \
MIFPA-11-02\ \ \ \ }
\end{flushright}

\vspace{25pt}
\begin{center}
{\large {\bf Critical Gravity in Four Dimensions}}

\vspace{15pt}

H. L\"u\hoch{1,2} and C.N. Pope\hoch{3,4}

\vspace{10pt}

\hoch{1}{\it China Economics and Management Academy\\
Central University of Finance and Economics, Beijing 100081, China}

\vspace{10pt}

\hoch{2}{\it Institute for Advanced Study, Shenzhen
University\\ Nanhai Ave 3688, Shenzhen 518060, China}

\vspace{10pt}

\hoch{3} {\it George P. \& Cynthia Woods Mitchell  Institute
for Fundamental Physics and Astronomy,\\
Texas A\&M University, College Station, TX 77843, USA}

\vspace{10pt}

\hoch{4}{\it DAMTP, Centre for Mathematical Sciences,
 Cambridge University,\\  Wilberforce Road, Cambridge CB3 OWA, UK}

\vspace{40pt}

\underline{ABSTRACT}
\end{center}

   We study four-dimensional gravity theories that are rendered
renormalisable by the inclusion of curvature-squared terms to the
usual Einstein action with cosmological constant.  By choosing the
parameters appropriately, the massive scalar mode can be eliminated
and the massive spin-2 mode can become massless.  This ``critical''
theory may be viewed as a four-dimensional analogue of chiral
topologically massive gravity, or of critical ``New Massive Gravity'' with a 
cosmological constant, in three dimensions.  We find that the
on-shell energy for the remaining massless gravitons vanishes. There
are also logarithmic spin-2 modes, which have positive energy.  The
mass and entropy of standard Schwarzschild type 
black holes vanish.  The critical theory might provide a 
consistent toy model for quantum gravity in four
dimensions.

\vspace{15pt}

\thispagestyle{empty}





\newpage

   Three-dimensional topologically massive gravity (TMG) \cite{tmg1}
has been studied extensively in recent times as a toy model for a
quantum theory of gravity. In addition to the usual massless
graviton (which carries no local degrees of freedom), there is in 
general a massive propagating spin-2 field.
For generic values of the coupling constant $\mu^{-1}$ for the
topological Chern-Simons term, the energy of massive spin-2
excitations is negative if one takes the Einstein-Hilbert term
to have the conventional (positive) sign.  On the other hand, the
BTZ black hole has positive mass if the Einstein-Hilbert term has
the conventional sign. Thus there is no choice of sign for the
Einstein-Hilbert term which gives positivity for both the massive
spin-2 excitations and the mass of the BTZ black hole solution.  It was,
however, noted in \cite{strom} that if the Chern-Simons coupling is
chosen so that $\mu\ell=1$, where $\ell$ is the ``radius'' of the
AdS$_3$ solution, then the massive spin-2 field becomes massless,
and both the field excitations and the BTZ black hole mass will be
positive for the conventional choice of sign for the
Einstein-Hilbert term.  It was, furthermore, conjectured that the
excitations in this ``critical'' theory are described by a consistent  
chiral two-dimensional boundary theory.

   In this paper, we address the question of whether any kind of
analogous critical limit might arise in four-dimensional gravity.
Even though such a limit would presumably not be expected to
describe a realistic theory of four-dimensional gravity, it might
nevertheless be of interest as another simplified ``toy model,''
with the advantage in this case of being in four, rather than three,
dimensions.  Since one would hope that such a toy model would be
renormalisable, the natural place to look is in the class of
four-dimensional gravity theories with curvature-squared
modifications, which were first studied in detail in
\cite{stelle1,stelle2}.  Since the Gauss-Bonnet invariant does not
contribute to the equations of motion in four dimensions, we just
need to consider the action
\be
I = \fft{1}{2\kappa^2}\,
  \int \sqrt{-g}\,d^4x(R -2\Lambda + \alpha R^{\mu\nu} R_{\mu\nu} +
   \beta R^2)\,.\label{action}
\ee

   As was discussed in \cite{stelle1,stelle2} (in the absence of the
cosmological term), this theory is renormalisable, and it describes
in general a massless spin-2 graviton, a massive spin-2 field and a
massive scalar.  The energies of excitations of the massive spin-2
field are negative, while those of the massless graviton are, as
usual, positive.  Thus although the theory is renormalisable, it
suffers from having ghosts.  The massive spin-0 is absent in the
special case $\alpha=-3\beta$, whilst the massive spin-2 is absent
if instead $\alpha=0$.

   Following the general strategy of \cite{strom}, we shall seek a limit
in which the massive spin-2 field becomes massless.  The presence of
the cosmological constant in the action (\ref{action}) is essential
for this step.  We shall also choose the parameters so that the
massive scalar is absent.  The energies of
excitations of the remaining massless graviton then vanish.  The 
fourth-order graviton operator also admits logarithmic modes, and we
find that these have positive energy.  We then
investigate the mass of black holes in the critical theory. We find
that the effect of the curvature-squared terms is to modify the mass
formula such that Schwarzschild-AdS black holes are massless. In
fact three-dimensional ``new massive gravity'' \cite{nmg} with a
cosmological constant exhibits similar
features at its critical point, 
of having zero energy excitations for the massless
gravitons \cite{liusun}, and zero mass for the BTZ black hole
\cite{hybrid2}.

   It is, of course, rather unusual to have a theory of gravity in
which black holes are massless.  One maybe has to accept this as the
price to be paid for having a four-dimensional theory of gravity,
without ghosts, that has the possibility of being renormalisable.
Thus although the theory cannot claim to be in any way
phenomenologically realistic, it may provide a useful simplified
arena for studying some aspects of a potentially renormalisable
theory of massless spin-2 fields in four dimensions.

  The equations of motion that follow from the action (\ref{action}) are
\bea
 && \cG_{\mu\nu} + E_{\mu\nu}=0\,,\label{eom}\\
\cG_{\mu\nu} &=& R_{\mu\nu} -\ft12 R\, g_{\mu\nu} + \Lambda\, g_{\mu\nu}\,,
  \label{cGdef}\\
E_{\mu\nu} &=& 2\alpha(R_{\mu\rho}\, R_\nu{}^\rho -\ft14 R^{\rho\sigma}
  R_{\rho\sigma}\, g_{\mu\nu}) + 2\beta R\, (R_{\mu\nu} -\ft14 R\, g_{\mu\nu})
\nn\\
&&+\alpha\, ( \square R_{\mu\nu} + \ft12 \square R\, 
  g_{\mu\nu} - 2\nabla_\rho \nabla_{(\mu} R_{\nu)}{}^\rho) +
 2\beta\, (g_{\mu\nu}\, \square R -\nabla_\mu\nabla_\nu R)\,.\label{Edef}
\eea
In what follows, we shall need to consider the linearisation of these
equations around a background solution of the equations of motion.  We shall
take the background to be four-dimensional anti-de Sitter spacetime, for
which, following from (\ref{eom}), we shall have
\be
R_{\mu\nu}=\Lambda\, g_{\mu\nu}\,,\qquad R= 4\Lambda\,,\qquad
R_{\mu\nu\rho\sigma}= \fft{\Lambda}{3}\, (g_{\mu\rho} g_{\nu\sigma} -
  g_{\mu\sigma} g_{\nu\rho})\,.\label{AdS4}
\ee
Note that in four dimensions, unlike in higher dimensions, the inclusion of
the explicit cosmological constant in (\ref{action}) is essential in order
to have an AdS solution. This is because $E_{\mu\nu}$ vanishes in any
Einstein space background in four dimensions. 

   Writing the varied metric as $g_{\mu\nu}\rightarrow g_{\mu\nu} +
h_{\mu\nu}$, and so $\delta g_{\mu\nu}=h_{\mu\nu}$, we find to first order
in variations that
\bea
\delta(\cG_{\mu\nu} + E_{\mu\nu})&=
  &[1+2\Lambda(\alpha+4\beta)]\, \cG^L_{\mu\nu}
+ \alpha\, \Big[ (\square -\fft{2\Lambda}{3}) \cG^L_{\mu\nu} -
\fft{2\Lambda}{3}\, R^L\, g_{\mu\nu}\Big]\nn\\
&&+ (\alpha+2\beta)\, [-\nabla_\mu\nabla_\nu + g_{\mu\nu}\, \square +
   \Lambda\, g_{\mu\nu}]R^L\,,\label{deom}
\eea
where $\cG^L_{\mu\nu}$ and $R^L$ are the linearised variations of
$\cG_{\mu\nu}$ and $R$:
\bea
\cG_{\mu\nu}^L &=& R^L_{\mu\nu} -\ft12 R^L\, g_{\mu\nu} -\Lambda\,
h_{\mu\nu}
\,,\label{GL}\\
R^L_{\mu\nu} &=& \nabla^\lambda\nabla_{(\mu} h_{\nu)\,\lambda}
   -\ft12\square h_{\mu\nu} -\ft12 \nabla_\mu\nabla_\nu h\,,\label{RicL}\\
R^L&=& \nabla^\mu\nabla^\nu h_{\mu\nu} -\square h - \Lambda h\,.\label{RL}
\eea
(We have also defined $R^L_{\mu\nu}$, the linearisation of $R_{\mu\nu}$,
and introduced $h=g^{\mu\nu} h_{\mu\nu}$.)

   For our purposes, it will be convenient to use general coordinate
invariance to impose the gauge condition
\be
\nabla^\mu h_{\mu\nu}= \nabla_\nu h\,.\label{gauge}
\ee
Substituting this into (\ref{GL})--(\ref{RL}), we find
\be
\cG^L_{\mu\nu} = -\ft12\square h_{\mu\nu} +\ft12 \nabla_\mu\nabla_\nu h
  +\fft{\Lambda}{3}\,h_{\mu\nu} +\fft{\Lambda}{6}\, h\,,\qquad
R^L = -\Lambda\, h\,.\label{GRgauge}
\ee
We can substitute these expressions into the variation of the
equations of motion, given by (\ref{deom}).  It is helpful first to
look at the trace, for which we find
\bea
0=g^{\mu\nu}\, \delta(\cG_{\mu\nu} + E_{\mu\nu}) =
  \Lambda\, [h - 2(\alpha+3\beta)\, \square h]\,.\label{traceeqn}
\eea
We see that $h$ describes a propagating massive scalar mode, except in the
special case that
\be
\alpha=-3\beta\,,\label{alphabeta}
\ee
in which case the equations of motion imply that $h=0$. It is this case,
where (\ref{alphabeta}) holds, that we shall focus on in our
subsequent discussion.\footnote{The condition (\ref{alphabeta}) has reproduced
the condition found in \cite{stelle2}, which had $\Lambda=0$, for the
absence of the massive scalar mode. Note that the gauge condition
(\ref{gauge}) would be rather degenerate if one started with $\Lambda=0$, since
the equation of motion for $h$ itself would not be independently determined.
The de Donder gauge would be a better choice in the theory with $\Lambda=0$.
Interestingly, however, we reproduce the correct massive scalar equation
from (\ref{traceeqn}), even for the $\Lambda=0$ theory, if we start with
$\Lambda\ne0$ and then take the limit where $\Lambda$ vanishes.}
  Note that modulo the Gauss-Bonnet combination, which does not contribute to
the equations of motion in four dimensions, the curvature-squared terms
with $\alpha=-3\beta$ can be written as $\ft12\alpha C^{\mu\nu\rho\sigma}
C_{\mu\nu\rho\sigma}$, where $C_{\mu\nu\rho\sigma}$ is the Weyl tensor.

Having imposed (\ref{alphabeta}), and hence determined that $h=0$, we are left
with the result that the variation of the field equations gives
\be
0=\delta(\cG_{\mu\nu}+E_{\mu\nu}) =
  \fft{3\beta}{2}\,\Big(\square -\fft{2\Lambda}{3}\Big)
\Big(\square -\fft{4\Lambda}{3} -\fft1{3\beta}\Big)\, h_{\mu\nu}\,,
\label{hTTeom}
\ee
where $h_{\mu\nu}$ is in the transverse traceless gauge
\be
\nabla^\mu\, h_{\mu\nu}=0\,,\qquad g^{\mu\nu}\, h_{\mu\nu}=0\,.\label{TTgauge}
\ee

   The fourth-order equation (\ref{hTTeom}) describes
a massless graviton,
satisfying
\be
\Big(\square -\fft{2\Lambda}{3}\Big)\, h^{(m)}_{\mu\nu}=0\,,\label{massless}
\ee
and a massive spin-2 field, satisfying
\be
\Big(\square -\fft{4\Lambda}{3} -\fft1{3\beta}\Big)\, h^{(M)}_{\mu\nu}=0\,.
\label{massive}
\ee
The criterion of stability for spin-2 modes satisfying
$(\square -2\Lambda/3  -M^2)h_{\mu\nu}=0$ in the AdS$_4$ background requires
that $M^2\ge0$ (see, for example, \cite{kkrep}), and so, since $\Lambda$ is
negative, we must have
\be 0<\beta \le \Big(-\fft1{2\Lambda}\Big)\,.\label{betarange} \ee
(In particular, $\beta$ must be positive.)
We are interested in choosing $\beta$ so that we are at the critical point
where the massive spin-2 field also becomes massless.  This is achieved
by taking
\be
\beta = -\fft{1}{2\Lambda}\,.\label{betacrit}
\ee

  By imposing (\ref{alphabeta}) in order to eliminate the massive scalar
mode, and additionally (\ref{betacrit}) in order to make the massive spin-2
mode become massless, we have arrived at a four-dimensional
theory describing only massless gravitons.  We may now examine the energy
of the excitations of the graviton modes in the AdS$_4$ background, 
which we take to be
\be
ds_4^2 = \fft{3}{(-\Lambda)}\, \Big[ -\cosh^2\rho \, dt^2 + d\rho^2
+\sinh^2\rho\, d\Omega_2^2\Big]\,.
\ee
A procedure for doing this
has been described in \cite{strom}, based upon the construction
of the Hamiltonian for the graviton field.  Leaving $\beta$ as yet
unrestricted, we may write down the quadratic action whose variation yields
the equations of motion (\ref{hTTeom}):
\bea
I_2 &=& -\fft{1}{2\kappa^2} \,
\int\sqrt{-g}\, d^4x\,h^{\mu\nu}(\delta\cG_{\mu\nu} +
   \delta E_{\mu\nu})\\
&=& -\fft{1}{2\kappa^2}\, \int \sqrt{-g}\, d^4x\, \Big[
\ft12(1+6\beta\Lambda)(\nabla^\lambda h^{\mu\nu})(\nabla_\lambda h_{\mu\nu})
  + \ft32\beta (\square h^{\mu\nu})(\square h_{\mu\nu})\nn\\
&&\qquad\qquad\qquad\qquad +
\fft{\Lambda}{3}\,(1+4\beta\Lambda) h^{\mu\nu}h_{\mu\nu}\Big]\,.\nn
\eea
Using the method of Ostrogradsky for Lagrangians written in terms of
second, as well as first, time derivatives we define the conjugate
``momenta''
\bea
\pi^{(1)\mu\nu} &=& \fft{\delta L_2}{\delta \dot h_{\mu\nu}} -
  \nabla_0 \Big(\fft{\delta L_2}{\delta(d(\nabla_0 h_{\mu\nu})/dt)}\Big)=
  -\fft{1}{2\kappa^2}\,
    \sqrt{-g}\,\nabla^0\Big( (1+6\beta\Lambda) h^{\mu\nu}
-3\beta \square h^{\mu\nu}\Big)\,,\nn\\
\pi^{(2)\mu\nu} &=& \fft{\delta L_2}{\delta (d(\nabla_0 h_{\mu\nu})/dt)}=
 -\fft{3\beta}{2\kappa^2}\, \sqrt{-g}\, g^{00}\,
   \square h^{\mu\nu}\,.
\eea
Since the Lagrangian is time-independent, the Hamiltonian is equal to its
time average, and writing it in this way 
is advantageous because we can then integrate time 
derivatives by parts.  Thus we obtain the Hamiltonian
\bea
H &=& T^{-1}\Big(\int d^4x\, \Big[\pi^{(1)\mu\nu}\, \dot h_{\mu\nu}
+\pi^{(2)\mu\nu}\, \fft{\del(\nabla_0 h_{\mu\nu})}{\del t} \Big] -I_2\Big)\nn\\
&=& \fft{1}{2\kappa^2T}\,
   \int \sqrt{-g}\, d^4x\, \Big[ -(1+6\beta\Lambda)\, \nabla^0 h^{\mu\nu}\,
       \dot h_{\mu\nu} +
              6\beta\, \Big(\fft{\del}{\del t} (\square h^{\mu\nu})\Big)
               (\nabla^0 h_{\mu\nu}) \Big] -\fft1{T}\, I_2\,,\label{H1}
\eea
where all time integrations are taken over the interval $T$.

  Evaluating this for the massless mode (satisfying (\ref{massless})),
and for the massive mode (satisfying (\ref{massive})), we therefore
obtain the on-shell energies
\bea E_{\rm massless} &=& -\fft{1}{2\kappa^2T}\,
(1+2\beta\Lambda)\,
   \int\sqrt{-g}\, d^4x\, \nabla^0 h^{\mu\nu}_{(m)}\,
   \dot h^{(m)}_{\mu\nu}\,,
\label{masslessE}\\
E_{\rm massive} &=& \fft{1}{2\kappa^2T}\,(1+ 2\beta\Lambda)\,
   \int\sqrt{-g}\, d^4x\, \nabla^0 h_{(M)}^{\mu\nu}\,
   \dot h^{(M)}_{\mu\nu}\,.
\label{massiveE}
\eea
Evidently, since the graviton modes in pure Einstein gravity with
$\beta=0$ are known to have positive energy, the integral itself in
(\ref{masslessE}) must be negative.  The integral in
(\ref{massiveE}) is therefore also expected to be negative (at least
when $\beta$ is chosen so that the mass of the mode is small, and
probably in all cases), and so we see that the massive excitations
in AdS, with positive $\beta$, will have negative energy. Imposing
now our criticality condition (\ref{betacrit}), we see that the
energies (\ref{masslessE}) and (\ref{massiveE}) become equal and
vanish. This is analogous to the critical situation \cite{liusun} of
new massive gravity \cite{nmg} with a cosmological constant.

   There are also logarithmic modes at the critical point, which are 
annihilated by the full fourth-order operator $(\square + \ft23\Lambda)^2$
but not by $(\square + \ft23\Lambda)$ alone.  (Analogous modes in 
three-dimensional chiral gravity were obtained in \cite{grujoh}.) 
Such logarithmic
modes have been constructed 
recently in \cite{behoroto}, where they were shown to have the form
$h_{\mu\nu}^{\rm log}= (2i t + \log\sinh 2\rho -\log\tanh\rho)\, h_{\mu\nu}$,
where $h_{\mu\nu}$ are standard spin-2 massless modes.  (See also 
\cite{alifar}.) We have explicitly
evaluated the expression (\ref{H1}) at the critical point for these
modes, and found, by numerical integration, that they have a finite and 
strictly positive energy.

   We now investigate the mass of black hole
solutions. (All solutions of the $\alpha=\beta=0$ theory are
also solutions of the full theory.) This can be done by
applying the procedure of Abbott and Deser
\cite{abodes} for calculating the mass of an asymptotically AdS
solution.  One does this by writing the black-hole metric in the
form $g_{\mu\nu}=\bar g_{\mu\nu} + h_{\mu\nu}$, where $\bar
g_{\mu\nu}$ is the metric on AdS, and interpreting the linearised
variation of the field equation, given in our case by (\ref{deom}),
as an effective gravitational energy-momentum tensor $T_{\mu\nu}$
for the black hole field. One then writes the conserved current
$J^\mu=T^{\mu\nu}\,\xi_\mu$, where $\xi^\mu$ is a Killing vector
that is timelike at infinity, as the divergence of a 2-form ${\cal
F}_{\mu\nu}$; {\it i.e.} $J^\mu= \nabla_\nu {\cal F}^{\mu\nu}$. From
this, one obtains the Abbott-Deser mass
\be
E =\fft{1}{2\kappa^2}\, \int_{S_\infty}\, dS_i\, {\cal F}^{0i}\,,
\label{AD}
\ee
as an integral over the sphere at infinity. The relevant
contributions to ${\cal F}^{\mu\nu}$ associated with the various
terms in (\ref{deom}) have been calculated in
\cite{destek}.  One may verify that defining
\bea
{\cal F}_{(0)}^{\mu\nu} &=& \xi_\alpha \nabla^{[\mu}
h^{\nu]\alpha}
 + \xi^{[\mu}\nabla^{\nu]}\, h + h^{\alpha[\mu}\nabla^{\nu]}\xi_\alpha
 -\xi^{[\mu}\, \nabla_\alpha h^{\nu]\alpha} + \ft12 h \nabla^\mu\xi^\nu\,,
 \nn\\
{\cal F}_{(1)}^{\mu\nu} &=& 2\xi^{[\mu}\, \nabla^{\nu]}\, R^L +
                                 R^L\, \nabla^\mu\xi^\nu\,,\nn\\
{\cal F}_{(2)}^{\mu\nu} &=& -2\xi_\alpha\, \nabla^{[\mu}\,
\cG_L^{\nu]\alpha}
 -2 \cG_L^{\alpha[\mu}\ \nabla^{\nu]}\, \xi_\alpha\,,
\eea
it follows that
\bea
\nabla_\nu\, {\cal F}_{(0)}^{\mu\nu} &=& \cG_L^{\mu\nu}\, \xi_\nu\,,\nn\\
\nabla_\nu\, {\cal F}_{(1)}^{\mu\nu} &=&\Big[(-\nabla_\mu\nabla_\nu +
   g^{\mu\nu}\square + \Lambda\, g^{\mu\nu}) R^L\Big] \xi_\nu\,,\nn\\
\nabla_\nu\, {\cal F}_{(2)}^{\mu\nu} &=&
  \Big[(\square - \fft{2\Lambda}{3})\cG_L^{\mu\nu} -
   \fft{2\Lambda}{3}\, R^L\, g^{\mu\nu}\Big]\, \xi_\nu\,.
\eea
These are precisely the three structures that we grouped together in
(\ref{deom}), after contracting $\delta(\cG_{\mu\nu} + E_{\mu\nu})$
with $\xi^\nu$.  It therefore becomes a mechanical exercise to assemble
the total tensor ${\cal F}^{\mu\nu}$, and then to evaluate the Abbott-Deser
mass for a black hole solution.

  Carrying out this procedure for the Schwarzschild-AdS black
hole, one finds that only the term in (\ref{AD}) coming
from the antisymmetric tensor ${\cal F}_{(0)}^{\mu\nu}$ gives a
non-vanishing contribution, and therefore the Abbott-Deser mass of
the Schwarzschild-AdS black hole is \cite{destek0}
\be M= m\, [1+2\Lambda(\alpha+4\beta)]= m\, (1 +
2\beta \Lambda)\,,\label{SdSmass} \ee
where $m$ is the usual mass parameter of the solution. Thus the
black hole mass is non-negative for $\beta$ in the range
(\ref{betarange}). For our critical condition (\ref{betacrit}) that
makes the massive spin-2 mode become massless, we see that
(\ref{SdSmass}) becomes
\be
M=0\,,
\ee
and so the Schwarzschild-AdS black hole has zero mass.  (We expect
the same to be true of Kerr-AdS black holes.) Similar zero-energy
results have previously been obtained in the context of a
scale-invariant gravity theory with a pure Weyl-squared action
\cite{bohost}, and in three-dimensional critical ``new massive gravity''
\cite{hybrid2}. (See also \cite{cacaoh}.)

Using the Wald \cite{wald} formula $S=-2\pi\int \sqrt{h}\,d^2x \,
\epsilon_{\alpha\beta}\epsilon_{\gamma\delta}\, (\del {\cal L}/\del
R_{\alpha\beta\gamma\delta})$ for the entropy of a black hole, we find
\be
S= \pi\,r_+^2\, [1+2\Lambda(\alpha+4\beta)]
\ee
for the Schwarzschild-AdS black hole, which is consistent with the
first law of thermodynamics $dM=T dS$ for all $\alpha$ and $\beta$,
including the critical case $\alpha=-3\beta=3/(2\Lambda)$ where both
$M$ and $S$ vanish.

   It was shown in \cite{stelle1} that four-dimensional Einstein
gravity with curvature-squared terms added is in general
renormalisable.  The case $\alpha=-3\beta$ was, however, excluded,
on the grounds that the scalar mode would then have a
propagator with only $1/k^2$ fall-off rather than $1/k^4$.  A key
new feature in our discussion is the inclusion of a cosmological
constant in the theory, leading, as we saw from equation
(\ref{traceeqn}), to the entire elimination of the scalar mode when
$\alpha=-3\beta$.  Thus, whether the critical theory we have
constructed might in fact be renormalisable now depends on whether
$\alpha=-3\beta$ and $\beta=-1/(2\Lambda)$ are stable under the 
renormalisation group
flow, which is an open question at this point.

In this paper, we have studied a four-dimensional theory of gravity
with cosmological constant, possibly rendered renormalisable by
including curvature-squared terms in the action. By choosing
parameters appropriately, we eliminated the massive scalar mode that
is generically present, and we also arranged for the massive spin-2
mode to become massless.  The resulting ``critical'' theory could be
viewed as a four-dimensional analogue of the chiral
three-dimensional TMG theory studied in \cite{strom}. It turns out
that the on-shell energy of the remaining massless spin-2 field
vanishes, while logarithmic modes of the fourth-order operator
have positive energy.  Furthermore, a calculation of the mass of
Schwarzschild-AdS black holes, based on the methods of Abbott, Deser
and Tekin, show that they are in fact massless in this theory. The
Schwarzschild-AdS black hole entropy is also zero in the critical theory.

   The fact that logarithmic modes of the fourth-order graviton operator 
have positive energy suggests
that despite being somewhat degenerate, the critical theory is not 
entirely trivial.  Although the standard Schwarzschild-AdS black hole has
zero mass and (consistently) zero entropy, there could in principle
exist more general black-hole solutions that are not also solutions
of pure Einstein gravity. It would be of considerable interest to try
to construct these, and to see whether they have non-zero mass.
The theory of critical gravity may provide an interesting toy model for
studying aspects of four-dimensional gravity and quantum gravity.

\section*{Acknowledgments}

We are grateful to the KITPC, Beijing, for hospitality during the
course of this work.  We thank Stanley Deser, 
Gary Gibbons, Haishan Liu, Yi Pang,
Malcolm Perry, Kelly Stelle, Bayram Tekin and Zhaolong Wang for 
helpful discussions.  The
research of C.N.P. is supported in part by DOE grant
DE-FG03-95ER40917.


\begin{thebibliography}{99}

\bibitem{tmg1} S. Deser, R. Jackiw and S. Templeton,
{\it Topologically massive gauge theories}, Annals Phys. {\bf 140},
372 (1982);
S. Deser, {\it Cosmological topological supergravity}, in
Christensen, S.M. (Ed.): Quantum Theory Of Gravity, 374-381 (1982).

\bibitem{strom} W. Li, W. Song and A. Strominger,
{\it Chiral gravity in three dimensions},
JHEP {\bf 0804}, 082 (2008), arXiv:0801.4566 [hep-th].

\bibitem{stelle1} K.S. Stelle,
{\it Renormalization of higher derivative quantum gravity},
Phys. Rev. {\bf D16}, 953 (1977).

\bibitem{stelle2} K.S. Stelle,
{\it Classical gravity with higher derivatives},
Gen. Rel. Grav. {\bf 9}, 353 (1978).

\bibitem{nmg} E.A. Bergshoeff, O. Hohm and P.K. Townsend,
{\it Massive gravity in three dimensions,}
  Phys. Rev. Lett.  {\bf 102}, 201301 (2009), arXiv:0901.1766 [hep-th].

\bibitem{liusun} Y. Liu and Y.W. Sun,
{\it Note on new massive gravity in AdS$_3$,}
  JHEP {\bf 0904}, 106 (2009), arXiv:0903.0536 [hep-th].


\bibitem{hybrid2}
  H. L\"u and Y. Pang,
{\it On hybrid (topologically) massive supergravity in three
dimensions,} arXiv:1011.6212 [hep-th].

\bibitem{grujoh}
D. Grumiller and N. Johansson,
{\it Instability in cosmological topologically massive gravity at the chiral
point},
JHEP {\bf 0807}, 134 (2008),
arXiv:0805.2610 [hep-th].


\bibitem{kkrep} M.J. Duff, B.E.W. Nilsson and C.N. Pope,
{\it Kaluza-Klein supergravity},
Phys. Rept. {\bf 130}, 1 (1986).

\bibitem{behoroto} E.A. Bergshoeff, O. Hohm, J. Rosseel and P.K. Townsend,
{\it Modes of log gravity},
arXiv:1102.4091 [hep-th].

\bibitem{alifar}M. Alishahiha and R. Fareghbal,
{\it $D$-dimensional log gravity},
arXiv:1101.5891 [hep-th].


\bibitem{abodes} L.F. Abbott and S. Deser,
{\it Stability of gravity with a cosmological constant},
Nucl. Phys. {\bf B195}, 76 (1982).

\bibitem{destek} S. Deser and B. Tekin,
{\it Energy in generic higher curvature gravity theories},
Phys. Rev. {\bf D67}, 084009 (2003), hep-th/0212292.

\bibitem{destek0} S. Deser and B. Tekin,
{\it Gravitational energy in quadratic curvature gravities},
Phys. Rev. Lett. {\bf 89}, 101101 (2002), hep-th/0205318.

\bibitem{bohost} D.G. Boulware, G.T. Horowitz and A. Strominger,
{\it Zero energy theorem for scale invariant gravity},
Phys. Rev. Lett. {\bf 50}, 1726 (1983).

\bibitem{cacaoh} R.G. Cai, L.M. Cao and N. Ohta,
{\it Black holes without mass and entropy in Lovelock gravity},
Phys. Rev. {\bf D81}, 024018 (2010), arXiv:0911.0245 [hep-th].

\bibitem{wald} R.M. Wald,
{\it Black hole entropy is the Noether charge},
Phys. Rev. {\bf D48}, 3427 (1993), gr-qc/9307038.



\end{thebibliography}
\end{document}